\documentclass[11pt, reqno]{amsart}
\usepackage{graphics,amsmath,amssymb,epsfig,stmaryrd,color}
\usepackage{amsfonts}
\usepackage{hyperref}

\hypersetup{
    colorlinks=true,
    urlcolor=blue,
    linkcolor=blue,
    citecolor=blue
}

\newtheorem{theorem}{Theorem}
\theoremstyle{plain}

\newtheorem{axiom}{Axiom}

\newtheorem{definition}{Definition}

\newtheorem{proposition}{Proposition}

\numberwithin{equation}{section}

\begin{document}
\title[Dual theory of choice]{Dual theory of choice with multivariate risks
(1)}
\author{Alfred Galichon}
\author{Marc Henry$^1$}

\begin{abstract}
We propose a multivariate extension of Yaari's dual theory of choice under risk. We show that a decision maker with a preference relation on multidimensional prospects that preserves first order stochastic dominance and satisfies comonotonic independence behaves as if evaluating prospects using a weighted sum of quantiles. Both the notions of quantiles and of comonotonicity are extended to the multivariate framework using optimal transportation maps. Finally, risk averse decision makers are characterized within this framework and their local utility functions are derived. Applications to the measurement of multi-attribute inequality are also discussed.
\end{abstract}

\maketitle



\noindent


{\footnotesize \textbf{JEL subject classification}: D63, D81, C61 \vskip5pt }

\section*{Introduction\protect\footnotetext[1]{%
This version: October 7, 2011. Correspondence address: Alfred Galichon, D\'{e}partement d'\'{e}conomie, \'{E}cole
polytechnique, 91128 Palaiseau, France and Marc Henry, D\'{e}partement de sciences \'{e}%
conomiques, Universit\'{e} de Montr\'{e}al, C.P. 6128, succursale
Centre-ville, Montr\'{e}al QC H3C 3J7, Canada. E-mail:
alfred.galichon@polytechnique.edu and marc.henry@umontreal.ca. Both authors
gratefully acknowledge support from the Chaire Axa \textquotedblleft
Assurance des Risques Majeurs\textquotedblright\ and the Chaire Soci\'{e}t%
\'{e} G\'{e}n\'{e}rale \textquotedblleft Risques
Financiers\textquotedblright . Galichon's research is partly supported by
the Chaire EDF-Calyon \textquotedblleft Finance and D\'{e}veloppement
Durable\textquotedblright\ and FiME, Laboratoire de Finance des March\'{e}s
de l'Energie (www.fime-lab.org). The authors thank Thibault Gajdos, John
Weymark and the participants at the 2010 ParisTech-\emph{Journal of Economic
Theory} Symposium on Inequality and Risk for helpful discussions and
comments. The authors are grateful to two anonymous referees and
particularly an Associate Editor for their careful reading of the manuscript
and insightful suggestions.}}

In his seminal paper \cite{Yaari:87}, Menahem Yaari proposed a theory of
choice under risk, which he called \textquotedblleft dual theory of
choice,\textquotedblright\ where risky prospects are evaluated using a
weighted sum of quantiles. The resulting utility is less vulnerable to
paradoxes such as Allais' celebrated paradox \cite{Allais:53}. The main
ingredients in Yaari's representation are the preservation of first order
stochastic dominance and insensitivity to hedging of comonotonic prospects.
Both properties have strong normative and behavioral appeal once it is
accepted that decision makers care only about the distribution of risky
prospects. The preservation of stochastic dominance is justified by the fact
that decision makers prefer risky prospects that yield higher values in all
states of the world, whereas comonotonicity captures the decision maker's
insensitivity to hedging comonotonic prospects, that is to say, the fact
that the decision maker who is indifferent between two prospects that yield
their higher and lower returns in the same states of the world, is also
indifferent between any convex combination of those prospects. The dual
theory has been used extensively as an alternative to expected utility in a
large number of contexts. The main drawback of the dual theory is that it
does not properly handle the case in which the prospects of consumptions of
different natures are not perfect substitutes. The assumption of \emph{law
invariance} of the decision functional (called neutrality in \cite{Yaari:87}
and by which the decision maker is insensitive to relabelings of the states
of the world) is easier to substantiate when several dimensions of the risk
are considered in the decision functional.

To handle these situations, we need to be able to express utility derived
from monetary consumption in different \emph{num\'{e}raires}, which is
easily done with Expected Utility Theory, but so far not covered by Yaari's
dual theory. Indeed, the latter applies only to risky prospects defined as
univariate random variables, thereby ruling out choice among
multidimensional prospects which are not perfect substitutes for each other,
such as risks involving both a liquidity and a price risk, collection of
payments in different currencies, payments at different dates, prospects
involving different goods of different natures such as consumption and
environmental quality, etc. Yaari \cite{Yaari:86} proposes a multivariate
version of his dual theory, but it involves independence of the risk
components and an axiom of separability (Axiom A in \cite{Yaari:86}), which
essentially removes the multidimensional nature of the problem.

We propose to remove this constraint with a multivariate extension of the
dual theory to risky prospects defined as random vectors that is applicable
as such to the examples listed above. The main challenge in this
generalization is the definition of quantile functions and comonotonicity in
the multivariate setting. Another challenge is to preserve the simplicity of
the functional representing preferences, so that they can be parameterized
and can be computed as efficiently as in the univariate case. Both
challenges are met with an appeal to optimal transportation maps that allow
for the definition of \textquotedblleft generalized
quantiles,\textquotedblright\ their efficient computation, and the extension
of comonotonicity as a notion of distribution free perfect correlation.
There are many ways of extending the notion of comonotonicity to a
multivariate framework consistently with the univariate definition. Our
proposed extension has the added property of preserving the equivalence
between comonotonicity and Pareto efficiency of allocations (see \cite%
{Landsberger} for the original result and \cite{CDG:09} for the multivariate
extension). With these notions of quantiles and comonotonicity in hand, we
give a representation of a comonotonic independent preference relation as a
weighted sum of generalized quantiles. The main difference between the
univariate case and the multivariate case is that comonotonicity and
generalized quantiles are defined with respect to an objective reference
distribution, which features in the representation. The reference
distribution is shown to be equal 
to the distribution of equilibrium prices in an economy with at least one
risk averse Yaari decision maker.

We then turn to the representation of a risk averse decision makers's
preferences within this theory. Risk aversion is defined in the usual way as
a preference for less risky prospects, where the notion of increasing risk
is suitably generalized to multivariate risky prospects. We show, again in a
direct generalization of the univariate case, that risk aversion is
characterized by a special form of the quantile weights defined above: risk
averse decision makers give more weight to low outcomes (low quantiles) and
less weight to high outcomes (high quantiles). As a result, given the
reference distribution with respect to which comonotonicity is defined, risk
averse decision makers are characterized by further simple restrictions on
their utility functionals, which makes this model as simple and as tractable
as expected utility. A further advantage of our decision functional is the
simple characterization of the local utility function and its close relation
to the multivariate quantile function.

The risk averse Yaari decision functional is a version of the Weymark social
evaluation function (in \cite{Weymark:81}) with a continuous state space.
Indeed, the formal equivalence between the evaluation of risky prospects and
the measurement of inequality noted in \cite{Atkinson:70} and \cite{Kolm:77}
allows us to draw implications of our theory for the measurement of
inequality of allocations of multiple attributes, such as consumption,
education, environment quality, etc. Seen as a social evaluation function,
our decision functional provides a compromise between the approach of \cite%
{GajdosWeymark} and \cite{Tsui:99} in that it allows a flexible attitude to
correlations between attributes without necessarily imposing correlation
aversion and thereby circumventing the Bourguignon-Chakravarty \cite{BC:2003}
critique of the assumption that attributes are substitutes rather than
complements.

The paper is organized as follows. The next section gives a short exposition
of the dual theory. The following section develops the generalized notion of
comonotonicity that is necessary for the multivariate extension, which is
given in Section~\ref{section: multivariate representation}. Risk aversion
is characterized in Section~\ref{section: risk aversion}. The economic
interpretation of the reference measure is given in Section~\ref{section:
economic interpretation} and the application to multi-attribute inequality
measurement is discussed in Section~\ref{section: inequality}. The final
section concludes.

\subsection*{Notation and basic definitions}

Let $({S},{\mathcal{F}},\mathbb{P})$ be a non-atomic probability space. Let $%
X:{S}\rightarrow {\mathbb{R}}^{d}$ be a random vector. We denote the
probability distribution of $X$ by ${\mathcal{L}}_{X}$. $\mathbb{E}$ is the
expectation operator with respect to $\mathbb{P}$. For $x$ and $y$ in ${%
\mathbb{R}}^{d}$, let $x\cdot y$ be the standard scalar product of $x$ and $y
$, and $\left\Vert x\right\Vert ^{2}$ the Euclidian norm of $x$. We denote
by $X=_{d}\mathcal{L}_{X}$ the fact that the distribution of $X$ is $%
\mathcal{L}_{X}$ and by $X=_{d}Y$ the fact that $X$ and $Y$ have the same
distribution. The \emph{equidistribution class} of $X=_{d}\mathcal{L}_{X}$,
denoted indifferently $\mbox{equi}(\mathcal{L}_{X})$ or $\mbox{equi}(X)$, is
the set of random vectors with distribution with respect to $\mathbb{P}$
equal to ${\mathcal{L}}_{X}$ (reference to $\mathbb{P}$ will be implicit
unless stated otherwise). $F_{X}$ denotes the cumulative distribution
function of distribution $\mathcal{L}_{X}$. $Q_{X}$ denotes its quantile
function. In dimension 1, this is defined for all $t\in \lbrack 0,1]$ by $%
Q_{X}(t)=\inf_{x\in \mathbb{R}}\{\mathrm{Pr}(X\leq x)>t\}$. In larger
dimensions, it is defined in Definition~\ref{definition: generalized
quantiles} of Section~\ref{subsection: multivariate quantiles} below. We
call $L_{d}^{2}$ the set of random vectors $X$ in dimension $d$ such that $%
\mathbb{E}\left[ \left\Vert X\right\Vert ^{2}\right] <+\infty $. We denote
by $\mathcal{D}$ the subset of $L_{d}^{2}$ containing random vectors with a
density relative to Lebesgue measure. A functional $\Phi $ on $L_{d}^{2}$ is
called upper semi-continuous (denoted u.s.c.) if for any real number $\alpha
$, $\{X\in L_{d}^{2}:\;\Phi (X)>\alpha \}$ is open. A functional $\Phi $ is
lower semi-continuous (l.s.c.) if $-\Phi $ is upper semi-continuous. For a
convex lower semi-continuous function $V:\mathbb{R}^{d}\mapsto \mathbb{R}$,
we denote by $\nabla V$ its gradient (equal to the vector of partial
derivatives). A doubly stochastic matrix is a square matrix of nonnegative
real numbers, each of whose rows and columns sum to $1$.

\section{Dual theory of decision under risk}

\label{section: dual theory} In this section, we first revisit Yaari's
\textquotedblleft Dual theory of choice under risk\textquotedblright\
presented in the eponymous paper \cite{Yaari:87}. As in \cite{Yaari:87}, we
consider a problem of choice among risky prospects as modeled by random
variables defined on an underlying probability space. The risky prospect $X$
is interpreted as a gamble or a lottery that a decision maker might consider
holding and the realizations of $X$ are interpreted as payments.

\subsection{Representation}

We suppose that the decision maker is characterized by a preference relation
$\succsim $ on the set of risky prospects. $X\succsim Y$ indicates that the
decision maker prefers prospect $X$ to prospect $Y$, $X\succ Y$ stands for $%
X\succsim Y$ and not $Y\succsim X$, whereas $X\sim Y$ stands for $X\succsim
Y $ and $Y\succsim X$. We first introduce the set of axioms satisfied by the
preference relation that were proposed by Yaari in \cite{Yaari:87}.

With the first axiom (which corresponds to Axioms~A2 and~A3 in Yaari \cite%
{Yaari:87}), we take the standard notion of preference as a continuous
pre-order (reflexive and transitive binary relation) which is complete.
Continuity of the preference relation is required relative to the topology
of weak convergence: a sequence of random prospects $X_n$ converges weakly
to $X$ if $\mathbb{E}f(X_n)$ converges to $\mathbb{E}f(X)$ for all
continuous bounded functions $f$ on $\mathbb{R}^d$. Then, $\succsim $ can be
represented by a continuous real valued function $\gamma $ in the sense that
$X\succsim Y$ if and only if $\gamma (X)\geq \gamma (Y)$.

\begin{axiom}
The preference relation $\succsim $ is reflexive, transitive, complete and
continuous relative to the topology of weak convergence. \label{axiom: weak
order}
\end{axiom}

A prospect $X$ is said to first order stochastically dominate a prospect $Y$
if there exist $\tilde{X}=_dX$ and $\tilde{Y}=_dY$ such that $X(s)\geq Y(s)$
for almost all states of the world $s\in S$. The following axiom requires
that whenever one prospect first order stochastically dominates a second,
then the former is preferred to the latter. This is formally stated as
follows.

\begin{axiom}
The preference $\succsim$ preserves first order stochastic dominance in the
sense that if prospect $X$ first order stochastically dominates prospect $Y$%
, then $X\succsim Y$, and if $X$ strictly first order stochastically
dominates prospect $Y$, then $X\succ Y$. \label{axiom: monotonicity}
\end{axiom}

Two prospects with the same distribution first order stochastically dominate
one another. Hence, Axiom~\ref{axiom: monotonicity} implies law invariance
of the preference relation, or what \cite{Yaari:87} calls \emph{neutrality},
i.e., $X=_{d}Y$ implies $X\sim Y$. Neutrality can be interpreted as the fact
that the decision maker is indifferent to relabelings of the states of the
world. Once neutrality is accepted, then Axiom~\ref{axiom: monotonicity} is
reasonable as it is equivalent to requiring that the decision maker prefers
prospects that yield a higher value in every state of the world. We shall
see below that with suitable extensions of the concepts of monotonicity and
stochastic dominance, this axiom remains reasonable in the multivariate
extension of Yaari's representation theorem.

Finally, the third axiom is the crucial one in this framework, as it
replaces independence by comonotonic independence. Recall that $X$ and $Y$
are comonotonic if $(X(s)-X(s^{\prime }))(Y(s)-Y(s^{\prime }))\geq 0$ for
all $s,s^{\prime }\in S$. The absence of a hedging opportunity between
comonotonic prospects justifies the requirement below.

\begin{axiom}
If $X,Y$ and $Z$ are pairwise comonotonic prospects, then for any $\alpha\in[%
0,1]$, $X\succsim Y$ implies $\alpha X+(1-\alpha)Z\succsim\alpha
Y+(1-\alpha)Z$. \label{axiom: comonotonic independence}
\end{axiom}

We can now state Yaari's representation result.

\begin{proposition}[Yaari]
A preference $\succsim $ on $[0,1]$-valued prospects satisfies Axioms~\ref%
{axiom: weak order}-\ref{axiom: comonotonic independence} if and only if
there exists a continuous non-decreasing function $f$ defined on $[0,1]$,
such that $X\succsim Y$ if and only if $\gamma (X)\geq \gamma (Y)$, where $%
\gamma $ is defined for all $X$ as $\gamma (X)=\int_{0}^{1}f(1-F_{X}(t))dt$. %
\label{proposition: Yaari}
\end{proposition}

This result is interpretable in terms of weighting of outcomes (through the
weighting of quantiles). Assume that each of the functions that we consider
satisfy the invertibility and regularity conditions needed to perform the
following operations. By integration by parts
\begin{equation*}
\int_{0}^{1}f(1-F_{X}(t))dt=\int_{0}^{1}f(1-u)dQ_{X}(u)=\int_{0}^{1}f(1-u)%
\frac{d}{du}Q_{X}(u)du=\int_{0}^{1}f^{\prime }(1-u)Q_{X}(u)du.
\label{YaariUnidim}
\end{equation*}%
Hence, calling $\phi (u)=f^{\prime }(1-u)$, we have the representation of $%
\succsim $ with the functional $\int_{0}^{1}\phi (t)Q_{X}(t)dt$. Hence,
increasing $f$ corresponds to positive $\phi $, which can be interpreted as
a weighting of the quantiles of the prospect $X$. As noted in \cite{Yaari:87}%
, the functional $\gamma $ satisfies $\gamma (\gamma (X))=\gamma (X)$, so
that $\gamma (X)$ is the certainty equivalent of $X$ for the decision maker
characterized by $\succsim $.

\subsection{Risk aversion}

We now turn to the characterization of risk averse decision makers among
those satisfying Axioms~\ref{axiom: weak order}-\ref{axiom: comonotonic
independence}. We define increasing risk as in Rothschild and Stiglitz \cite%
{RS:70}. The formulations in the first part of the definition below are
equivalent by the Blackwell-Sherman-Stein Theorem (see, for instance,
Chapter~7 of \cite{SS:2007}).

\begin{definition}[Concave ordering, risk aversion]
a) A prospect $Y$ is dominated by $X$ in the \emph{concave ordering},
denoted $Y\leq_{cv} X$, when the equivalent statements (i) or (ii) hold:

\begin{itemize}
\item[(i)] for all continuous concave functions, $\mathbb{E}f(Y)\leq \mathbb{%
E}f(X)$.

\item[(ii)] $Y$ has the same distribution as $\hat{Y}$ where $(X,\hat{Y})$
is a martingale, i.e., $\mathbb{E}(\hat{Y}|X)=X$ ($\hat{Y}$ is sometimes
called a \emph{mean-preserving spread} of $X$).
\end{itemize}

\label{definition: increasing risk} b) The preference relation $\succsim$ is
called \emph{risk averse} if $X\succsim Y$ whenever $X\geq_{cv} Y$. \label%
{definition: risk aversion}
\end{definition}

Notice that $x\rightarrow x$ and $x\rightarrow -x$ are both continuous and
concave function. Therefore, condition (i) in the first part of the
definition implies that $\mathbb{E}[X]=\mathbb{E}[Y]$ is necessary for a
concave ordering relationship between $X$ and $Y$ to exist. With this
definition, we can recall the characterization of risk averse preferences
satisfying Axioms~\ref{axiom: weak order}-\ref{axiom: comonotonic
independence} as those with convex $f$ (see Section~5 of \cite{Yaari:87} or
Theorem~3.A.7 of \cite{SS:2007}).

\begin{proposition}
A preference relation $\succsim $ satisfying Axioms~\ref{axiom: weak order}-%
\ref{axiom: comonotonic independence} is risk averse if and only if the
function $f$ in Theorem~\ref{proposition: Yaari} is convex.\label%
{proposition: univariate risk averse representation}
\end{proposition}

This monotonicity of the derivative of $f$ has the natural interpretation
that risk averse decision makers evaluate prospects by giving high weights
to low quantiles (corresponding to low values of the prospect) and low
weights to high quantiles. Indeed, with the formulation $\gamma \left(
X\right) =\int_{0}^{1}\phi (u)Q_{X}(u)du$ and the identification $\phi
(u)=f^{\prime }(1-u)$, an increasing convex $f$ corresponds to positive
decreasing $\phi $, and therefore to a weighting scheme in which low
quantiles (corresponding to unfavorable outcomes) receive high weights and
high quantiles (corresponding to favorable outcomes) receive low weights.

\section{Multivariate quantiles and comonotonicity}

\label{section: multivariate quantiles and comonotonicity}

The main ingredients in our multivariate representation theorem are the
multivariate extensions of quantiles and comonotonicity. As we shall see,
the two are intimately related.

\subsection{Multivariate quantiles}

\label{subsection: multivariate quantiles}

We first note that the quantile of a random variable can be characterized as
an increasing rearrangement of the latter. Hence, by classical rearrangement
inequalities, quantiles are solutions to maximum correlation problems. More
precisely, by the rearrangement inequality of Hardy, Littlewood and P\'{o}%
lya \cite{HLP52}, we have the following well known equality:
\begin{equation}
\int_{0}^{1}tQ_{X}(t)d(t)=\max \left\{ \mathbb{E}[XU]:\;U%
\mbox{
uniformly distributed on }[0,1]\right\} ,  \label{equation: polya}
\end{equation}%
where the quantile function $Q_{X}$ has been defined above. This variational
characterization is crucial when generalizing Yaari's representation theorem
to the multivariate setting. Indeed, consider now a random vector $X$ on $%
\mathbb{R}^{d}$ and a reference distribution $\mu $ on $\mathbb{R}^{d}$,
with $U$ distributed according to $\mu $. We introduce \emph{maximum
correlation functionals} to generalize the variational formulation of (\ref%
{equation: polya}).

\begin{definition}[Maximal correlation functionals]
A functional $\varrho _{\mu }:L_{d}^2\rightarrow \mathbb{R}$ is called a
maximal correlation functional with respect to a reference distribution $\mu$
if for all \ $X\in L_{d}^2$,
\begin{equation*}
\varrho _{\mu }(X):= \sup \left\{ \mathbb{E}[X\cdot \tilde{U}]:\;\tilde{U}%
=_d\mu \right\}.
\end{equation*}%
\label{definition: maximal correlation functional}
\end{definition}

It follows from the theory of optimal transportation (see Theorem 2.12(ii),
p. 66 of \cite{Villani:2003}) that if $\mu $ is absolutely continuous with
respect to Lebesgue measure (which will be assumed throughout the rest of
the paper), then there exists a convex lower semi-continuous function $V:{%
\mathbb{R}}^{d}\rightarrow {\mathbb{R}}$ and a random vector $U$ distributed
according to $\mu $ such that $X=\nabla V(U)$ holds $\mu $-almost surely,
and such that $\varrho _{\mu }(X)=\mathbb{E}[\nabla V(U)\cdot U]$. In that
case, the pair $\left( U,X\right) $ is said to achieve the \emph{optimal
quadratic coupling of }$\mu $ with respect to the distribution of $X$. The
function $V$ is called the \emph{transportation potential} of $X$ with
respect to $\mu $ or the \emph{transportation potential} from $\mu $ to the
probability distribution of $X$.\footnote{$V$ is convex and hence
differentiable except on set of measure zero by Rademacher's Theorem
(Theorem 2.4 in \cite{Preiss:90}), so that the expression $\mathbb{E}[\nabla
V(U)\cdot U]$ above is well defined.} This shows that the gradient of the
convex function $V$ thus obtained satisfies the multivariate analogue of
equation~(\ref{equation: polya}). We therefore adopt $\nabla V$ as our
notion of a generalized quantile.

\begin{definition}[$\protect\mu $-quantile]
The $\mu $-quantile function of a random vector $X$ on $\mathbb{R}^{d}$ with
respect to an absolutely continuous distribution $\mu $ on $\mathbb{R}^{d}$
is defined by $Q_{X}=\nabla V$, where $V$ is the transportation potential of
$X $ with respect to $\mu $. \label{definition: generalized quantiles}
\end{definition}

This concept of a multivariate quantile is the counterpart of our definition
of multivariate comonotonicity in the representation theorem, and the
latter, introduced in the following section has strong economic
underpinnings, as discussed in Section~\ref{section: economic interpretation}%
, where we give the economic interpretation of the reference measure $\mu$.

\subsection{Multivariate comonotonicity}

\label{subsection: multivariate comonotonicity}

Two univariate prospects $X$ and $Y$ are comonotonic if there is a prospect $%
U$ and non-decreasing maps $T_{X}$ and $T_{Y}$ such that $Y=T_{Y}(U)$ and $%
X=T_{X}(U)$ almost surely or, equivalently, $\mathbb{E}[UX]=\max \left\{
\mathbb{E}[\tilde{U}X]:\;\tilde{U}=_{d}U\right\} $ and $\mathbb{E}[UY]=\max
\left\{ \mathbb{E}[\tilde{U}Y]:\;\tilde{U}=_{d}U\right\} $. Comonotonicity
is hence characterized by maximal correlation between the prospects over the
equidistribution class. This variational characterization (where products
will be replaced by scalar products) will be the basis for our generalized
notion of comonotonicity.

\begin{definition}[$\protect\mu$-comonotonicity]
Let $\mu$ be a probability measure on $\mathbb{R}^d$ with finite second
moments. A collection of random vectors $X_i\in L_d^2$, $i\in I$, are called
$\mu $-comonotonic if one has
\begin{equation*}
\varrho_\mu\left(\sum_{i\in I} X_i\right)=\sum_{i\in I}{\varrho_\mu( X_i)}.
\end{equation*}
\label{definition: mu comonotonicity}
\end{definition}

When $\mu $ is absolutely continuous with respect to Lebesgue measure, it
follows from the representation of $\varrho _{\mu }$ that the family $X_{i}$
is $\mu $-comonotonic if and only if there exists a vector $U$ distributed
according to $\mu $ such that $U\in \mbox{argmax}_{\tilde{U}}\{\mathbb{E}%
[X_{i}\cdot \tilde{U}],\;\tilde{U}=_{d}\mu \}$ for all $i\in I$. In other
words, the $X_{i}$'s can be rearranged simultaneously so that they achieve
maximal correlation with $U$. When the distributions of the random vectors
are absolutely continuous with respect to Lebesgue measure, the concept of $%
\mu $-comonotonicity is transitive.

\begin{proposition}
Suppose that $X$ and $Y$ are $\mu$-comonotonic and that $Y$ and $Z$ are $\mu
$-comonotonic, with the distribution of $Y$ assumed to be absolutely
continuous with respect to Lebesgue measure. Then $X$ and $Z$ are $\mu$%
-comonotonic. \label{proposition: transitivity}
\end{proposition}

\subsubsection*{Comonotonic allocations and Pareto efficiency}

It is worth discussing this definition of comonotonicity as generalizations
of the classical univariate notion of comonotonicity are not unique. The
main motivation for introducing it is to generalize the univariate
equivalence between comonotonic and Pareto efficient allocations in a
risk-sharing economy. Consider an Arrow-Debreu economy with $n$ agents, and
with an aggregate endowment which is a random vector $X$. Thus the $i$-th
dimension of the realization $X^{i}\left( \omega \right) $ in state $\omega
\in \Omega$ of this random vector is the quantity of good $i\in \left\{
1,...,d\right\} $ produced in this state. An \emph{allocation} (or \emph{%
risk-sharing allocation}) of $X$ is a sharing rule of this aggregate
endowment among the $n$ agents, hence it is the specification of $n$ random
vectors $X_{1},...,X_{n}$ such that%
\begin{equation*}
\forall \omega \in \Omega,~\sum_{k=1}^{n}X_{k}\left( \omega \right) =X\left(
\omega \right)
\end{equation*}%
where $X_{k}^{i}\left( \omega \right) $ is the quantity of good $i$
allocated to agent $k\in \left\{ 1,...,n\right\} $ in state $\omega $. An
allocation is called (Pareto) \emph{efficient} if no other allocation
dominates the former, agent by agent, in the sense of the concave ordering
(as defined in Proposition~\ref{proposition: concave-ordering} below).

In dimension one, it is known since the seminal paper of Landsberger and
Meilijson \cite{Landsberger} that a risk-sharing allocation is Pareto
efficient with respect to the concave order if and only if it is
comonotonic. That is, given any comonotonic allocation, it is not possible
to find another allocation such that each risky endowment would be preferred
under the new allocation by every risk-averse decision maker to the
endowments in the original allocation. Multivariate generalization of this
equivalence is not obvious, but it turns out that, as recently shown by
Carlier, Dana and Galichon \cite{CDG:09}, this result can be extended to the
multivariate case, with comonotonicity replaced by \emph{multivariate
comonotonicity}, if one defines an allocation to be comonotonic in the
multivariate sense if and only if it is $\mu $-comonotonic for some measure $%
\mu $ with enough regularity. In our view, this result strongly supports the
claim that our notion of comonotonicity is in some sense the
\textquotedblleft natural\textquotedblright\ multivariate extension of
comonotonicity.

\subsubsection*{Relation with other multivariate notions of comonotonicity}

Puccetti and Scarsini \cite{PS:08} have also applied the theory of optimal
transportation to generalize the notion of comonotonicity to the
multivariate setting. They review possible multivariate extensions of
comonotonicity, including the notion of $\mu$-comonotonicity that we
propose. But the concept they favor differs from ours in the sense that
according to their notion of multivariate comonotonicity (which they call
\emph{$c$-comonotonicity}), two vectors $X$ and $Y$ are $c$-comonotonic if
and only $(X,Y)$ is an optimal quadratic coupling. That is, $X$ and $Y$ are
c-comonotonic if and only if there is a convex function $V$ such that $%
Y=\nabla V(X)$ holds almost surely. However, unlike $\mu $-comonotonicity, $c
$-comonotonicity is in general not transitive, and does not seem to be
related to efficient risk-sharing allocations and equilibrium.

Schmeidler \cite{Schmeidler:89} introduces an internal notion of
comonotonicity: if a decision maker evaluates prospects according to $%
\succsim$, then Schmeidler-comonotonicity of two prospects $X$ and $Y$ means
that for all pairs of states of the world $(s,t)$, $X(s)\succsim X(t)$
implies $Y(s)\succsim Y(t)$, i.e., prospects $X$ and $Y$ are more desirable
in the same states of the world. In contrast, we extend the Weymark \cite%
{Weymark:81} - Yaari \cite{Yaari:87} motivation in our definition of
comonotonicity, which can be related to the state prices in the economy, as
explained in Section~\ref{section: economic interpretation}. The two notions
have no obvious relation, as we see by considering two $\mu$-comonotonic
prospects $X$ and $Y$ and imposing Schmeidler comonotonicity. By $\mu$%
-comonotonicity, there exists $U=_d\mu$ and generalized quantile functions $%
Q_X$ and $Q_Y$ such that $X=Q_X(U)$ and $Y=Q_Y(U)$.
Schmeidler-comonotonicity of $X$ and $Y$ would require that the univariate
random variables $(\mathbb{E}U)\cdot X$ and $(\mathbb{E}U)\cdot Y$ are
comonotonic in the usual sense. Although they are equivalent in dimension
one, in higher dimensions, neither of these two concepts implies the other.

\section{Multivariate Representation Theorem}

\label{section: multivariate representation} Now that we have given a
formalization of the notion of maximal correlation in a law invariant sense
that is suitable for a multivariate extension of Yaari's dual theory, we can
proceed to generalize Yaari's representation result to the case of a
preference relation among multivariate prospects. We consider prospects,
which are elements of $L_{d}^{2}$. Axiom~$1^{\prime}$ below is a mild
smoothness requirement for the preference relation. A functional $\gamma$ is
called Fr\'echet differentiable in $X$ relative to the $L_d^2$ metric if
there is a linear functional $L$ such that $|\gamma(X+h)-\gamma(X)-L(h)|/%
\sqrt{\mathbb{E}[h^2]}\rightarrow0$. As in \cite{CKS:87}, the functional
will not be Fr\'echet differentiable at all points; we only require
differentiability at one point.\vskip5pt\noindent \textbf{Axiom $1^{\prime }$%
. }The preference $\succsim $ is represented by a continuous functional $%
\gamma $ on $L_{d}^{2}$ such that at at least one point its Fr\'{e}chet
derivative exists and is non-zero.

Given sufficient regularity, first order stochastic dominance can be
characterized equivalently by pointwise dominance of cumulative distribution
functions or pointwise dominance of quantile functions. It is the latter
that we adopt for our multivariate definition.

\begin{definition}[$\protect\mu $-first order stochastic dominance]
A prospect $X$ $\mu$-first order stochastically dominates prospect $Y$
relative to the componentwise partial order $\geq $ on $\mathbb{R}^{d}$ if $%
Q_{X}(t)\geq Q_{Y}(t)$ for almost all $t\in {\mathbb{R}}^{d}$, where $Q_{X}$
and $Q_{Y}$ are the generalized quantiles of $X$ and $Y$ with respect to a
distribution $\mu $ on ${\mathbb{R}}^{d}$. \label{definition: first order
stochastic dominance}
\end{definition}

For any $U=_{d}\mu $, we have $Q_{X}(U)=_{d}X$ and $Q_{Y}(U)=_{d}Y$. If $X$ $%
\mu $-first order stochastically dominates $Y$, then $Q_{X}(U)\geq Q_{Y}(U)$
almost surely. Hence, $\hat{X}\geq \hat{Y}$ almost surely for some $\hat{X}%
=_{d}X$ and $\hat{Y}=_{d}Y$, which is the \textquotedblleft usual
multivariate stochastic order\textquotedblright\ (see \cite{SS:2007}, p.
266). The converse does not hold in general.

The remaining two axioms require fixing an absolutely continuous reference
probability distribution $\mu$ on $\mathbb{R}^d$. \vskip5pt\noindent \textbf{%
Axiom $2^\prime$. }The preference $\succsim$ preserves $\mu$-first order
stochastic dominance in the sense that if prospect $X$ $\mu$-first order
stochastically dominates prospect $Y$, then $X\succsim Y$, and if $X$ $\mu$%
-first order strictly stochastically dominates prospect $Y$, then $X\succ Y$%
. \vskip5pt The extension of the comonotonicity axiom is the key to the
generalization of the dual theory to multivariate prospects. The statement
of Axiom~\ref{axiom: comonotonic independence} is unchanged, but the concept
of comonotonicity is now dependent on a reference distribution $\mu $. The
prospects $X,\,Y$ and $Z$ are comonotonic, or more precisely $\mu $%
-comonotonic, if they are all \emph{maximally correlated} in the law
invariant sense of Definition~\ref{definition: mu comonotonicity} with a
reference $U$ (where $U$ has distribution $\mu $).

\vskip5pt\noindent \textbf{Axiom $3^\prime$. }If $X,\,Y$ and $Z$ are $\mu$%
-comonotonic prospects, then for any $\alpha\in[0,1]$, $X\succsim Y$ implies
$\alpha X+(1-\alpha)Z\succsim\alpha Y+(1-\alpha)Z$.\vskip5pt

We are now in a position to state the multivariate extension of Yaari's
representation theorem.

\begin{theorem}[Multivariate Representation]
A preference relation on multivariate prospects in $L_d^2$ satisfies Axioms~$%
1^{\prime }$, $2^{\prime }$ and $3^{\prime }$ relative to a reference
probability measure $\mu $ if and only if there exists a function $\phi $
such that for $U=_{d}\mu $, $\phi (U)\in L_d^2$, $\phi (U)\in (\mathbb{R}%
_{-})^{d}$ almost surely and such that for all pairs $X,Y$, $X\succsim Y$ if
and only if $\gamma (X)\geq \gamma (Y)$, where $\gamma $ is defined for all $%
X$ by $\gamma (X)=\mathbb{E}[Q_{X}(U)\cdot \phi (U)]$, where $Q_{X}$ is the $%
\mu $-quantile of $X$.\label{theorem: representation}
\end{theorem}

When $d=1$, the representation is independent of $\mu$ and we recover the
result of Proposition~\ref{proposition: Yaari}. As in the univariate case,
the decision maker assesses prospects with a weighting scheme $\phi$ of
quantiles of the prospects. Because $\gamma $ in Theorem~\ref{theorem:
representation} satisfies $\gamma(\gamma(X))=\gamma(X)$, $\gamma (X)$ is the
certainty equivalent of $X$ as in the univariate case. Furthermore, $%
\succsim $ satisfies \emph{linearity in payments}, i.e., for any positive
real number $a$ and any $b\in \mathbb{R}^{d}$ (identified with a constant
multivariate prospect), $\gamma (aX+b)=a\gamma (X)+b$.

It should be noted that Choquet expected utility \cite{Schmeidler:89}
handles multivariate prospects under Schmeidler comonotonicity (defined in
Section~\ref{subsection: multivariate comonotonicity}). As shown in \cite%
{Wakker:90}, under Axiom~\ref{axiom: monotonicity}, Choquet expected utility
is identical to the functional of Proposition~\ref{proposition: Yaari}.
Hence, when restricted to decision under risk, Choquet expected utility
aggregates the multiple dimensions of the prospects with the utility
function and then considers univariate quantiles of the resulting utility
index. This is in contrast with the functional of Theorem~\ref{theorem:
representation}, which directly evaluates multivariate quantiles of the
prospects and thereby models attitudes to substitution risk between the
dimensions of the prospect.

\section{Risk aversion and the local utility function}

\label{section: risk aversion} In this section, we consider the question of
representing those decision makers satisfying Axioms~$1^{\prime }$, $%
2^{\prime }$ and $3^{\prime }$ that are risk averse in the sense of
Definition~\ref{definition: risk aversion}. We then show that the \emph{%
local utility function} in the sense of \cite{Machina:1982} is easily
computable and provides an interpretation of the reference distribution $\mu$%
.

\subsection{Risk aversion}

For our characterization of risk averse Yaari decision makers, we need to
generalize the concept of a mean-preserving spread to the multivariate
setting.

\begin{proposition}[Concave ordering]
For any prospects $X$ and $Y$ whose respective distributions are absolutely
continuous with respect to Lebesgue measure, the following properties are
equivalent.

\begin{itemize}
\item[(a)] For every bounded concave function $f$ on $\mathbb{R}^{d}$, $%
\mathbb{E}f(X)\geq \mathbb{E}f(Y)$

\item[(b)] $Y=_d \hat{Y}$, with $\mathbb{E}[\hat{Y}\vert X]=X$.

\item[(c)] $\varrho_\mu(X)\leq \varrho_\mu(Y)$ for every probability measure
$\mu$.

\item[(d)] $X$ belongs to the closure of the convex hull of the
equidistribution class of $Y$.

\item[(e)] $\Phi (X)\geq \Phi (Y)$ for every u.s.c. law-invariant concave
functional $\Phi :L_{d}^{2}\rightarrow {\mathbb{R}}$.
\end{itemize}

\label{proposition: concave-ordering} When any of the properties above hold,
one says that $Y$ is dominated by $X$ in the \emph{concave ordering},
denoted $Y\leq_{cv} X$.
\end{proposition}

Statements (a) and (b) are identical in the multivariate case as in \cite%
{RS:70}. The equivalence between the two is a classical result that can be
traced back at least to \cite{Strassen:65} (see section~\ref{Proof of
concave ordering} for details). The interpretation of the ordering as a
preference ordering for all risk aversion expected utility maximizers (a)
and as an ordering of mean-preserving spreads (b) also carry over to the
multivariate case. Statement (d) is the continuous equivalent to
multiplication by a doubly stochastic matrix.

As in the univariate case and Definition~\ref{definition: increasing risk}%
(b), risk averse decision makers will be defined by aversion to
mean-preserving spreads. It turns out that imposing risk aversion on a
preference relation that satisfies axioms~$1^\prime$, $2^\prime$ and $%
3^\prime$ is equivalent to requiring the following property, sometimes
called \emph{preference for diversification}.

\begin{axiom}
For any two preference equivalent prospects $X$ and $Y$ (i.e., such that $%
X\sim Y$), convex combinations are preferred to either of the prospects,
(i.e., for any $\alpha \in \lbrack 0,1]$, $\alpha X+(1-\alpha )Y\succsim X$%
). \label{axiom: preference for diversification}
\end{axiom}

This is formalized in the following theorem, which gives a representation
for risk averse \emph{Yaari decision makers}.

\begin{theorem}
In dimension $d\geq 2$, for a preference relation satisfying Axioms $%
1^{\prime }$, $2^{\prime }$ and $3^{\prime }$, the following statements are
equivalent:

\begin{itemize}
\item[(a)] $\succsim$ is risk averse, namely $X\succsim Y$ whenever $%
X\geq_{cv} Y$.

\item[(b)] $\succsim $ satisfies Axiom~\ref{axiom: preference for
diversification}.

\item[(c)] The function $\phi$ involved in the representation of the
preference relation in Theorem~\ref{theorem: representation} satisfies $-
\phi(u)= \alpha u+u_0$ for $\alpha>0$ and $u_0\in\mathbb{R}^d$.
\end{itemize}

\label{theorem: risk averse representation}
\end{theorem}

So, in the multivariate setting the functional $\gamma $ is convex if and
only if $\phi (x)=-\alpha x-x_{0}$ for some $\alpha $ real positive and $%
x_{0}\in \mathbb{R}^{d}$. This is a major difference with dimension one,
where the functional is convex if and only if $-\phi $ is a non-decreasing
map. This implies that a multivariate Yaari risk averse decision maker is
entirely characterized by the reference distribution $\mu $.

\subsection{Local Utility Function}

Throughout the rest of the paper, we shall assume that the conditions in
Theorem \ref{theorem: risk averse representation} are met. Hence, our
discussion of local utility functions will be limited to the case of risk
averse decision makers. By law-invariance, we denote $\gamma (P):=\gamma (X)$%
, where $X=_{d}P$. Without loss of generality, we shall also assume that $%
\phi (u)=-u$, thus $\gamma (P):=-\mathbb{E}[\nabla V_{P}(U)\cdot U]$, where $%
V_{P}=V_{X}$ is the transportation potential (see section~\ref{subsection:
multivariate quantiles} for the definition) from the reference probability
distribution $\mu $ of $U$ to the probability distribution $P$ of $X$. As we
have seen, the gradient $\nabla V_{P}$ of this transportation potential is
the $\mu $-quantile function of distribution $P$.

As shown in \cite{Machina:1982}, when smoothness requirements are met, a
local analysis can be carried out in which a (risk-averse) non-Expected
Utility function behaves for small perturbations around a fixed risk in that
same way as a (concave) utility function. Formally, the local utility
function is defined as $u\left( x|P\right) =D_{P}\gamma (x)$, where $%
D_{P}\gamma $ is the Fr\'{e}chet derivative of $\gamma $ at $P$ (see Section~%
\ref{section: multivariate representation} for the definition). Denoting by $%
V^{\ast }(x)=\sup_{u}[u\cdot x-V(u)]$ the \emph{Legendre-Fenchel transform}
of a convex lower semicontinuous function $V$, we have:
\begin{eqnarray*}
\gamma \left( P\right) &=&\mathbb{E}[\nabla V_{P}(U)\cdot U]=\max \left\{
\mathbb{E}[X\cdot U]:\;X=_{d}P,\;U=_{d}\mu \right\} \\
&=&\min \left\{ \int V(u)d\mu (u)+\int V^{\ast }(x)dP(x):\;V\;%
\mbox{ convex and
l.s.c. }\right\} \\
&=&\int V_{P}(u)d\mu (u)+\int V_{P}^{\ast }(x)dP(x)
\end{eqnarray*}%
by the duality of optimal transportation (see, for instance, Theorem~2.9, p.
60 of \cite{Villani:2003}).

Defining $f(V,Q)=\int Vd\mu +\int V^{\ast }dQ$, we have $\gamma
(P)=-\inf_{V}f(V,P)$. Hence, an envelope theorem argument formally yields $%
u(x|P)=D_{P}\gamma (x)=-V_{P}^{\ast }(x)$. Therefore, \emph{the local
utility function is $-V_{P}^{\ast }$, the (negative of the) Legendre-Fenchel
transform of the transportation potential $V_{P}$}. This point sheds light
on the economic interpretation of this potential, thanks to Machina's theory
of local utility. The function $-V_{P}^{\ast }$ is concave, which is
consistent with the risk aversion of a Yaari decision maker given the
assumptions of Theorem~\ref{theorem: risk averse representation}. For
univariate prospects, $u(x|P)=-V_{P}^{\ast }(x)=\int_{-\infty }^{x}F_{X}(z)dz
$, so that we recover the fact that when $X=_{d}P$ is a mean-preserving
spread of $Y=_{d}Q$, $u(z|P)\leq u(z|Q)$ for all $z$.

\section{Economic interpretation of the reference measure \label{section:
economic interpretation}}

We now discuss the behavioral interpretation of $\mu $. As we saw in Theorem %
\ref{theorem: representation}, the generalization of the Yaari preferences
to the multivariate case led us to define a utility functional $\gamma $
over prospects such that $\gamma (X)=\mathbb{E}[X\cdot \phi (\tilde{U})]$
for some prospect $\tilde{U}$ which is correlated to $X$. $\tilde{U}$ is an
index such that $X\cdot \tilde{U}$ measures how favorable the outcome is for
the decision maker. $\phi (\tilde{U})$ is a weighting of the contingent
outcome $X$, so that $\gamma $ over- or under-weights prospects in each
state using weights $\phi (\tilde{U})$. Hence, the dispersion of $\mu $
induces a departure from risk-neutrality. In the special case in which $\mu $
is the distribution of a constant $u_{0}$, $\gamma (X)=\mathbb{E}[X\cdot
\phi (u_{0})]=\mathbb{E}[X]\cdot \phi (u_{0})$ and one recovers the case of
a risk-neutral decision maker. On the contrary, when $\mu $ exhibits
considerable dispersion, then the variance of $\phi (\tilde{U})$ is large in
general, so that the \textquotedblleft favorable\textquotedblright\ outcomes
(in the sense that $X\cdot \tilde{U}$ is high) are weighted less, at least
if $\phi (\tilde{U})=-\alpha \tilde{U}$. This induces risk aversion. When $%
\phi (\tilde{U})$ differs from a rescaling of $\tilde{U}$, there may be some
discrepancy between the weighting of a given state and how favorable it is.
Hence, the variance of $\mu $ is no longer directly associated with risk
aversion.

We now turn to the equilibrium implications of the reference measure $\mu $
and show how it is related to the distribution of the state prices in an
economy in equilibrium when a decision maker with risk averse decision
functional as in Theorem~\ref{theorem: risk averse representation} is
present in the economy. 
%
Consider an economy where one of the agents (whom we shall refer to as
\textquotedblleft Yaari\textquotedblright ) has preferences as in Theorem~%
\ref{theorem: risk averse representation}, with reference measure $\mu $.
Assume that there is a risk sharing equilibrium in this economy, which is
supported by the stochastic discount factor $\xi $, meaning that if the
original risky endowment of the agent is $X_{0}$, then the agent's budget
set is $\{X:\mathbb{E}[(X-X_{0})\cdot \xi ]=0\}$. The demand for risk $X$ of
Yaari is therefore $\max_{\hat{X}}\gamma (\hat{X})\ $subject to $\mathbb{E}[(%
\hat{X}-X_{0})\cdot \xi ]=0.$ Since Yaari is assumed risk-averse, $\gamma $
is concave, and the demand for risk $X$ satisfies the local optimality
condition $\max_{\hat{X}}\mathbb{E}[u(\hat{X}|P)]\ $subject to $\mathbb{E}[(%
\hat{X}-X_{0})\cdot \xi ]=0$. The first order conditions yield $\nabla
u(X|P)=\lambda \xi $, where $\lambda $ is the Lagrange multiplier associated
with the budget constraint, where $\lambda \neq 0$ unless there is no trade
in equilibrium. Now, as explained above, $u(x|P)=-V_{P}^{\ast }(x)$, hence $%
\nabla u(X|P)=-\nabla V_{P}^{\ast }(X)$. Now, by definition of the
transportation potential $V_{p}$ from $\mu $ to $P$, $\nabla V_{P}^{\ast
}(X)=_{d}\mu $. Hence $-\lambda \xi =_{d}\mu $ which implies that $\mu $ is
(up to scale) the distribution of the stochastic discount factor $\xi $.
Therefore, when there is a Yaari decision maker with reference measure $\mu $
in the economy, the stochastic discount factor should be distributed
according to $\mu $. This result is an extension of the well-known result
that states that when there is a risk-neutral decision maker in the economy,
the stochastic discount factor should equal one, that is, the risk-neutral
probability should coincide with the actuarial probability. To summarize, if
a risk-sharing equilibrium exists with a Yaari risk-averse decision maker
with reference measure $\mu $, then $\mu $ coincides with the distribution
of the stochastic discount factor. Thereby, $\mu $ is related to the
distribution of the state prices.

\section{Relation with multi-attribute inequality measurement}

\label{section: inequality} The theory developed here has implications for
inequality rankings of allocations of multiple attributes (such as income,
education, environmental quality, etc.) in a population. Atkinson \cite%
{Atkinson:70} recognized the relevance of stochastic orderings to the
measurement of inequality and its foundation on principles such as the
desirability of Pigou-Dalton transfers (also known as Pigou-Dalton
Majorization). Weymark \cite{Weymark:81} added to Pigou-Dalton Majorization
a principle of comonotonic independence, which he interpreted as neutrality
to the source of variation in income, and obtained a class of social
evaluation functions, which he called generalized Gini evaluation functions.
The functional form is identical to the decision functional derived
independently on a continuous state space by \cite{Yaari:87}. Indeed, \cite%
{Kolm:77} notes the formal equivalence between the problem of decision under
risk and the measurement of inequality. The random vector of risks or
prospects that we consider in the present work can be interpreted as an
allocation of multiple attributes over a continuum of individuals. With this
interpretation, states of the world are identified with individuals in the
population and the decision function $\gamma $ is interpreted as a social
evaluation function. Law invariance (Yaari neutrality, i.e., insensitivity
to relabelings of the states of the world) of the decision functional is
thus equivalent to anonymity of the social evaluation function. The ranking
of ordinally equivalent allocations obtained through Pigou-Dalton
Majorization (see\cite{Kolm:77}) corresponds to the concave ordering
discussed in Proposition~\ref{proposition: concave-ordering}. More
precisely, the mean-preserving spread characterization (b) in Proposition~%
\ref{proposition: concave-ordering} is equivalent to (d), which is the
infinite-dimensional analogue of multiplication by a doubly stochastic
matrix. Hence, our risk averse multivariate Yaari decision functional can be
interpreted as a social evaluation function for allocations of multiple
attributes, which satisfies anonymity, monotonicity and Pigou-Dalton
Majorization in the sense of Theorem~3 in \cite{Kolm:77}.

The inequality literature achieves functional forms for social evaluation
functions in the multi-attribute case by adding two distinct types of
majorization principles that allow the comparison of non-ordinally
equivalent social evaluations. Tsui \cite{Tsui:95}, \cite{Tsui:99} considers
correlation increasing transfers. Gajdos and Weymark \cite{GajdosWeymark}
extend generalized Gini social evaluation functions to the multivariate case
with a comonotonic independence axiom. Two allocations are said to be
comonotonic if all individuals are ranked identically in all attributes
(i.e., the richest is also the most educated etc.), and the ranking between
two comonotonic allocations is not reversed by the addition of a comonotonic
allocation. They use an attribute separability axiom (Axiom A in \cite%
{Yaari:86}) to reduce the dimensionality of the problem via independence of
the attributes. Specifically, Theorem~4 of \cite{GajdosWeymark} is a special
case of our Theorem~\ref{theorem: risk averse representation} when the
attribute vector $X$ and the reference distribution $\mu $ both have
independent marginals. Our representation can also incorporate trade-offs
between attributes and attitudes to correlations between attributes of the
kind that are entertained in \cite{Tsui:99}, but is not restricted to the
latter. Correlation aversion would correspond to perceived substitutability,
but perceived complementarity can also be entertained in our approach,
thereby circumventing Bourguignon and Chakravarty's critique of correlation
increasing majorization (in \cite{BC:2003}) based on the observation that
\textquotedblleft there is no a priori reason for a person to regard
attributes as substitutes only. Some of the attributes can as well be
complements\textquotedblright\ (p. 36).

\section{Conclusion}

We have developed concepts of quantiles and comonotonicity for multivariate
prospects, thus allowing for the consideration of choice among vectors of
payments in different currencies, at different times, in different
categories of goods, etc. The multivariate concepts of quantiles and
comonotonicity were used to generalize Yaari's dual theory of choice under
risk, where decision makers that are insensitive to hedging of comonotonic
risks are shown to evaluate prospects using a weighted sum of quantiles.
Risk averse decision makers were shown to be characterized within this
framework by a reference distribution, making the dual theory as readily
applicable as expected utility. Risk attitudes were also analyzed from the
point of view of a local utility function. Implications for the ranking of
increasing risk aversion is the topic of further research. Applications of
the representation theorem to the measurement of multi-attribute inequality
were also discussed. The flexibility in its handling of attitudes to
correlation between attributes is a promising feature of the decision
functional.

\appendix

\section{\protect\scriptsize Proof of results in the main text}

\subsection{{\protect\scriptsize Proof of Proposition~\protect\ref%
{proposition: transitivity}}}

{\scriptsize By definition, there are two convex lower semi-continuous
functions $V_{1}$ and $V_{2}$ and a random vector $U=_{d}\mu $ such that $%
X=\nabla V_{1}(U)$ and $Y=\nabla V_{2}(U)$ almost surely. Similarly, there
are convex functions $V_{3}$ and $V_{4}$ and a random vector $\tilde{U}$
such that $Y=\nabla V_{3}(\tilde{U})$ and $Z=\nabla V_{4}(\tilde{U})$. Now
the assumptions on the absolute continuity of $\mu $ and the distribution of
$Y$ imply that $\nabla V_{2}$ is essentially unique. Hence, $\nabla
V_{2}=\nabla V_{3}$ and, therefore, $U=\tilde{U}$ holds almost surely. It
follows that $X$ and $Z $ are $\mu $-comonotonic. \hskip4pt$\square $ }

\subsection{{\protect\scriptsize Proof of Proposition~\protect\ref%
{proposition: concave-ordering}}}

{\scriptsize \label{Proof of concave ordering} }

{\scriptsize The equivalence between (a) and (b) is a famous result stated
and extended by many authors, notably Hardy, Littlewood, P\'olya, Blackwell,
Stein, Sherman, Cartier, Fell, Meyer and Strassen. See Theorem~2 of \cite%
{Strassen:65} for an elegant proof. We now show that (b) implies (c).
Suppose (b) holds. As explained in Section 2.1, there exists a map $\zeta $
such that $\varrho _{\mu }(X)=\mathbb{E}[\zeta (X)\cdot X]$ and $\zeta
(X)=_{d}\mu $. Now, $\mathbb{E}[\zeta (X)\cdot X]=\mathbb{E}[\zeta (X)\cdot
\mathbb{E}[\hat{Y}|X]]=\mathbb{E}[\zeta (X)\cdot \hat{Y}],$ which is less
than $\varrho _{\mu }(Y).$ Next, we show that (c) implies (d). Indeed, the
convex closure $\overline{co}(equi(Y))$ of the equidistribution class of $Y$
is a closed convex set and hence characterized by its support functional $%
\varrho _{\mu }(Y)$. Therefore, $X\in \overline{co}(equi(Y))$ is equivalent
to $\mathbb{E}[Z\cdot X]\leq \varrho _{\mu }(Y)$ for all $Z$, which in turn
is equivalent to $\varrho _{\mu }(X)\leq \varrho _{\mu }(Y)$.\ Now, we show
that (d) implies (e). Indeed, if $X\in \overline{co}(equi(Y))$, then there
is a sequence $(Y_{k}^{n})_{k\leq n}$ of random vectors each distributed as $%
Y$ and positive weights $\alpha _{k}^{n}$ such that $\sum_{k=1}^{n}\alpha
_{k}^{n}=1$ and $X=\lim_{n\rightarrow \infty }\sum_{k=1}^{n}\alpha
_{k}^{n}Y_{k}^{n}$. Then, for any law invariant concave functional, we have $%
\Phi \left( \sum_{k=1}^{n}\alpha _{k}^{n}Y_{k}^{n}\right) \leq
\sum_{k=1}^{n}\alpha _{k}^{n}\Phi \left( Y_{k}^{n}\right) =\Phi (Y)$ and the
conclusion follows by upper semi-continuity. Finally, (e) implies (a)
because when $\mathcal{L}_{X}$ is absolutely continuous with respect to
Lebesgue measure, for any bounded concave function $f$, $X\mapsto \mathbb{E}%
f(X)$ is a law invariant concave upper semi-continuous functional. \hskip4pt$%
\square $ }

\subsection{{\protect\scriptsize Proof of Theorem~\protect\ref{theorem:
representation}}}

{\scriptsize \label{Proof of representation} }

{\scriptsize Note first that $\gamma $ defined for all prospects $X$ by $%
\gamma (X)=\mathbb{E}[Q_{X}(U)\cdot \phi (U)]$ for a function $\phi $ such
that $\phi (U)\in (\mathbb{R}_{-})^{d}$ is Lipschitz and monotonic, so that
Axioms~$1^{\prime }$ and~$2^{\prime }$ are satisfied for a preference
relation represented by $\gamma $. Finally, comonotonic independence follows
directly from the fact that for any two prospects $X$ and $Y$, the
generalized quantile functions $Q_{X}$, $Q_{Y}$ and $Q_{X+Y}$ satisfy $%
Q_{X+Y}(U)=Q_{X}(U)+Q_{Y}(U)$. We now show this fact. By the definition of
the generalized quantile functions, we have $\mathbb{E}[Q_{X+Y}(U)\cdot
U]=\sup_{\tilde{U}=_{d}U}\mathbb{E}[(X+Y)\cdot \tilde{U}]\leq \sup_{\tilde{U}%
=_{d}U}\mathbb{E}[X\cdot \tilde{U}]+\sup_{\tilde{U}=_{d}U}\mathbb{E}[Y\cdot
\tilde{U}]=\mathbb{E}[Q_{X}(U)\cdot U]+\mathbb{E}[Q_{Y}(U)\cdot U]$. On the
other hand, we also have $\mathbb{E}[Q_{X+Y}(U)\cdot U]=\sup_{\tilde{Z}%
=_{d}X+Y}\mathbb{E}[\tilde{Z}\cdot U]\geq \mathbb{E}[(Q_{X}(U)+Q_{Y}(U))%
\cdot U]$ since by construction, $Q_{X}(U)=_{d}X$ and $Q_{Y}(U)=_{d}Y$, and
the desired equality follows.}

{\scriptsize Conversely, we now prove that a preference relation $\succsim $
satisfying Axioms~$1^{\prime }$, $2^{\prime }$ and $3^{\prime }$ is
represented by a functional $\gamma $ defined for all prospect $X$ by $%
\gamma (X)=\mathbb{E}[Q_{X}(U)\cdot \phi (U)]$ for a function $\phi $ such
that $\phi (U)\in (\mathbb{R}_{-})^{d}$ almost surely. By Axiom~$1^{\prime }$%
, there exists a functional $\gamma $ representing $\succsim $ and there is
a point $Z\in L_{d}^{2}$, where $\gamma $ is Fr\'{e}chet differentiable with
non-zero gradient $D$. Let $Q_{Z}$ be the generalized quantile of $Z$
relative to $\mu $. There exists a $U\in L_{d}^{2}$ with distribution $\mu $
such that $Z=Q_{Z}(U)$ almost surely. Let $X$ and $Y$ be two prospects in $%
L_{d}^{d}$ with $\mu $-quantile functions $Q_{X}$ and $Q_{Y}$ respectively.
By the definition of $\mu $-comonotonicity, $Q_{X}(U)$, $Q_{Y}(U)$ and $%
Z=Q_{Z}(U)$ are $\mu $-comonotonic. By Axiom~$2^{\prime }$, $\gamma $ is law
invariant, so that $\gamma (X)\geq \gamma (Y)$ is equivalent to $\gamma
(Q_{X}(U))\geq \gamma (Q_{Y}(U))$. Hence, by Axiom~$3^{\prime }$, $\gamma
(X)\geq \gamma (Y)$ implies that for any $0<\epsilon \leq 1$, we have $%
\gamma (\epsilon Q_{X}(U)+(1-\epsilon )Z)\geq \gamma (\epsilon
Q_{Y}(U)+(1-\epsilon )Z)$. Hence, $\gamma (Z+\epsilon (Q_{X}(U)-Z))\geq
\gamma (Z+\epsilon (Q_{Y}(U)-Z))$ and, therefore, $\gamma (Z)+\mathbb{E}%
[D\cdot \epsilon (Q_{X}(U)-Z)]\geq \gamma (Z)+\mathbb{E}[D\cdot \epsilon
(Q_{Y}(U)-Z)]-o(\epsilon )$, or, finally, $\mathbb{E}[D\cdot Q_{X}(U)]\geq
\mathbb{E}[D\cdot Q_{Y}(U)]$.}

{\scriptsize Suppose now that $X$ and $Y$ are two prospects such that $%
\mathbb{E}[D\cdot Q_{X}(U)]=\mathbb{E}[D\cdot Q_{Y}(U)]$. We shall show that
$\gamma (Q_{X}(U))=\gamma (Q_{Y}(U))$ and, hence, that $\gamma (X)=\gamma (Y)
$, thereby concluding that the functional $X\mapsto \mathbb{E}[D\cdot
Q_{X}(U)]$ represents $\succsim $. Indeed, suppose that $\mathbb{E}[D\cdot
Q_{X}(U)]=\mathbb{E}[D\cdot Q_{Y}(U)]$. We will show shortly that there
exists a function $\phi $ such that $\mathbb{E}[D\cdot \nabla \phi (U)]>0$
and, hence, that $\mathbb{E}[D\cdot (Q_{X}(U)+\epsilon \nabla \phi (U))]>%
\mathbb{E}[D\cdot Q_{Y}(U)]$ and $\mathbb{E}[D\cdot (Q_{X}(U)-\epsilon
\nabla \phi (U))]<\mathbb{E}[D\cdot Q_{Y}(U)]$ for any $\epsilon >0$. Using
the result above yields $\gamma (Q_{X}(U)+\epsilon \nabla \phi (U))\geq
\gamma (Q_{Y}(U))$ and $\gamma (Q_{X}(U)-\epsilon \nabla \phi (U))\leq
\gamma (Q_{Y}(U))$. Hence, $\gamma (Q_{X}(U))=\gamma (Q_{Y}(U))$ by the
continuity of $\gamma $. Let us now show that $\mathbb{E}[D\cdot \nabla \phi
(U)]=0$ for all gradient functions $\nabla \phi $ yields a contradiction.
Calling $V_{Z}$ the convex function such that $Z=Q_{Z}(U)=\nabla V_{Z}(U)$
almost surely, $D$ is the Fr\'{e}chet derivative of $\gamma $ at $Z=\nabla
V_{Z}(U)$. Hence, $\mathbb{E}[D\cdot \nabla \phi (U)]=0$ implies that $%
\gamma (\nabla (V_{Z}(U)+\epsilon \phi (U)))=\gamma (\nabla
V_{Z}(U))+o(\epsilon )$. This is true for all gradient functions $\nabla
\phi $ and, in particular, for $\phi =(V_{\epsilon }-V_{Z})/\epsilon $,
where $V_{\epsilon }$ is such that $Z_{\epsilon }=\nabla V_{\epsilon }(U)$
converges to $Z$ in $L^{2}$. We then have $\gamma (Z_{\epsilon })-\gamma
(Z)=o(\epsilon )$ and, hence, $D=0$, which contradicts Axiom~$1^{\prime }$.
We have shown that $\succsim $ is represented by the functional $X\mapsto
\mathbb{E}[D\cdot Q_{X}(U)]$. As $\mu $ is absolutely continuous with
respect to Lebesgue measure, $D$ can be written as $\phi (U)$ for some
function $\phi $ which takes values in $(\mathbb{R}_{-})^{d}$ by Axiom~$%
2^{\prime }$. \hskip4pt$\square $ }

\subsection{{\protect\scriptsize Proof of Theorem~\protect\ref{theorem: risk
averse representation}}}

{\scriptsize That (a) implies (b) follows from Proposition~\ref{proposition:
concave-ordering}. We now show that (b) implies (c). Axiom 4 implies that $%
\gamma (Q_{X}+\alpha (\tilde{X}-Q_{X}))\geq \gamma (X)$ for all $\alpha \in
(0,1]$ and all $\tilde{X}$ in the equidistribution class of $X$. The
representation of Theorem~\ref{theorem: representation} implies the
differentiability of $\gamma $ at $Q_{X}(U)$ for any $X$, call $D_{X}$ its
gradient. This implies that $\mathbb{E}[\tilde{X}\cdot D_{X}]\leq \mathbb{E}%
[Q_{X}(U)\cdot D_{X}]$ for all $\tilde{X}=_{d}X$. Hence, $\gamma
(X)=-\varrho _{{\mathcal{L}}_{\phi (U)}}(X)$. Thus, by Axiom $3^{\prime }$,
comonotonicity with respect to $\mu $ implies comonotonicity with respect to
${\mathcal{L}}_{\phi (U)}$. By Lemma 10 in \cite{EGH:2009}, this implies
that $\phi (U)=-\alpha U-x_{0}$ for some $\alpha >0$ and $x_{0}\in \mathbb{R}%
^{d}$, and the result follows. Finally, we show that (c) implies (a). Assume
(c), in which case for all $X\in \mathbb{L}_{d}^{2}$, $-\gamma (X)=\alpha
\mathbb{E}[Q_{X}(U)\cdot U]+u_{0}\cdot \mathbb{E}[X]$. Thus, $-\gamma
(X)=\alpha \varrho _{\mu }(X)+u_{0}\cdot \mathbb{E}[X]$. Therefore, by
Proposition \ref{proposition: concave-ordering}, $X\geq _{cv}Y$ implies $%
\varrho _{\mu }(X)\leq \varrho _{\mu }(Y)$, and so $\gamma (X)\geq \gamma (Y)
$. \hskip4pt$\square $ }

\end{document}